# Cosmic Rays and Terrestrial Life: a Brief Review

Dimitra Atri[1,2] and Adrian L. Melott[3]
1. Department of High Energy Physics, Tata Institute of Fundamental Research, Colaba, Mumbai 400 005, India, atri.dimitra@gmail.com
2. Blue Marble Space Institute of Science, Seattle, WA 98145-1561, USA
3. Department of Physics and Astronomy, University of Kansas, Lawrence, KS 66045, USA, melott@ku.edu

**Abstract**:

"The investigation into the possible effects of cosmic rays on living organisms will also offer great interest." – Victor F. Hess, Nobel Lecture, December 12, 1936

High-energy radiation bursts are commonplace in our Universe. From nearby solar flares to distant gamma ray bursts, a variety of physical processes accelerate charged particles to a wide range of energies, which subsequently reach the Earth. Such particles contribute to a number of physical processes occurring in the Earth system. A large fraction of the energy of charged particles gets deposited in the atmosphere, ionizing the atmosphere, causing changes in its chemistry and affecting the global electric circuit. Remaining secondary particles contribute to the background dose of cosmic rays on the surface and parts of the subsurface region. Life has evolved over the past ~ 3 billion years in presence of this background radiation, which itself has varied considerably during the period [1-3]. As demonstrated by the Miller-Urey experiment, lightning plays a very important role in the formation of complex organic molecules, which are the building blocks of more complex structures forming life. There is growing evidence of increase in the lightning rate with increasing flux of charged particles. Is there a connection between enhanced rate of cosmic rays and the origin of life? Cosmic ray secondaries are also known to damage DNA and cause mutations, leading to cancer and other diseases. It is now possible to compute radiation doses from secondary particles, in particular muons and neutrons. Have the variations in cosmic ray flux affected the evolution of life on earth? We describe the mechanisms of cosmic rays affecting terrestrial life and review the potential implications of the variation of high-energy astrophysical radiation on the history of life on earth.

1. Introduction

### 1.1 Background of the possible production sources of cosmic rays: astrophysical sources, fluxes and their probabilities

The Sun is the primary source of radiation on Earth. The total solar irradiance (TSI) is eight orders of magnitude larger in energy flux than cosmic rays. Its photon spectrum peaks in the optical yellow region and the particle spectrum in KeV energies. The maximum energy of solar particles is ~MeV except in case of solar flares or CMEs [4], where the energy can reach a few 10s of GeV [5]. However, in order to have an effect on the biosphere, the radiation should be capable of significantly altering the atmospheric chemistry and/or generating secondary particles so that the radiation dose on the ground is increased significantly.

The geomagnetic field provides a good shield against ~ MeV particles, which therefore do not have any significant effect on the biosphere at large. However, since the magnetic field lines guide the charged particles towards the magnetic poles, even low energy particles can impact life in polarregions. Since the flux of Galactic Cosmic Rays (GCRs) is modulated by solar activity, events such as a Forbush decrease (when the solar wind sweeps more galactic cosmic rays away from the Earth) and geomagnetic storms can vary the flux of cosmic rays on short intervals too. Higher energy particles are generated in supernovae shocks, whose estimated rate is about 2-3 per century in our galaxy [6]. A powerful shock accelerates charged particles through the diffusive shock acceleration process. The estimated energy can go up to PeV ranges [7].

**1.2 Variation in CR rate: periodic and non-periodic variations**

Since charged particles are filtered by the solar and geomagnetic fields, a variation in such fields changes the flux of cosmic rays. A periodic variation in cosmic ray flux is anti-correlated with the solar magnetic field, and is well known as the 11-year solar cycle. The Earth's magnetic field is also known to undergo variations, also known as magnetic field reversals, with apparently total irregularity. However, it has been shown that changes in cosmic ray flux due to these reversals are too small to cause any substantial impact on the biosphere [8]. Other minor variations are caused by the periodic lunisolar cycles caused due to the disruption of the Earth-Sun magnetic link by the passage of the moon. Going to longer time periods, the Milankovitch cycles are caused due to the changes in the Earth's orbit around the sun in a periodic manner [9]. It has also been suggested that the Earth may be exposed to a periodic dose of high-energy cosmic rays due to our Galaxy's infall toward the Virgo cluster coupled with the oscillatory movement of our solar system perpendicular to the galactic plane [10]. This hypothesis predicts an enhanced exposure every ~62 Million years.

Non-periodic variation can occur due to proximity of the Earth with an intense radiation source, such as supernovae [11, 12]. Movement of our solar system into a dense interstellar cloud can push back the solar wind plasma, which protects the Earth from GCRs. With little shielding from this plasma, there will be an increase in the rate of Anomalous Cosmic Rays, which can cause severe ozone damage [13]. Large solar proton events occur at a higher rate than nearby supernovae, and can have a moderate impact on the biosphere [14].

## 2. Mechanisms of CR Interaction with the atmosphere

### 2.1. Ionization and atmospheric electricity

Cosmic ray primaries undergo hadronic interactions upon impact with the atmosphere. The resulting particles either decay or further interact, generating other particles. All charged particles produced in the atmosphere undergo electromagnetic interactions and thus form the 'electromagnetic component' of an air shower. If the energy of the particle is sufficient, it can knock down electrons while traversing through the atmosphere, thereby ionizing it. The location of the peak of atmospheric ionization depends on the primary energy [15]. For the normal GCR spectrum, the peak lies in the stratosphere. Higher energy primaries are capable of penetrating deeper and their peaks are further below, closer towards the Earth.

The mechanism of cosmic ray induced atmospheric ionization's role in generating thunderstorms is not yet established in quantitative terms and research in the field is underway. However the following mechanism is widely agreed upon: Atmospheric electric fields present in thunderclouds are not strong enough to initiate electric breakdown. However, electrons generated by air showers are energetic and can knock down more electrons, creating more ionization, and an electron avalanche is formed. Upon reaching the critical energy, electrons become relativistic runways and result in an abrupt discharge, releasing the energy in form of thunderstorms [16, 17]. The value of critical electric field is found to be $1.8\,\rho$ MeV/cm, where $\rho$ is the density in grams per cubic centimeter at a given altitude. This process becomes more efficient with primaries of higher energy as they generate more electrons and deposit more energy in the lower atmosphere. GCRs are also known to modulate the global electrical circuit The concentration of atmospheric ions directly changes with charged particle flux of GCRs and indirectly affects the charges in the troposphere though the modulation of current flow in the global electric circuit [18].

Electric fields of ~10 kV/m are typical in case of thunderstorms. A strong correlation is also observed between cosmic ray intensity and the magnitude of electric field disturbances in some experiments [19]. It must be mentioned that different types of cloud layers produce different types of thunderstorms, which vary in their magnitude and polarity. Both increase and decrease in the intensity in cosmic ray secondaries can thus be observed by experiments depending on the type of event. Work is under progress to get a quantitative understanding of thunderstorms and associated particle acceleration in the atmosphere.

The role of cosmic rays affecting the cloud cover and its impact on the climate [20] has been a topic of intense debate [21]. According to the hypothesis, an increase in comic ray intensity will increase the rate of ionization in the atmosphere and would result in increased cloud formation rate [20]. As the cloud cover would increase, less amount of radiation would reach the Earth's surface and would result in global cooling. Experimental work is underway in CERN to test this hypothesis [22]. If this hypothesis is true, it could explain the early faint young sun paradox. The early Sun is expected to have had ~30% less luminosity compared to the present value and one would expect extremely low temperatures as a result. However, geological records indicate the presence of liquid water on Earth during that era, which contradicts the above statement [23]. This paradox can be partially resolved when one considers cosmic ray flux during that period. Since the sun was more active ~ 4 Gyr ago, cosmic ray shielding would have been higher from intense solar winds. With lower flux of cosmic rays, the cloud cover is expected to be smaller and would result in 'global warming', thus explaining the paradox [24]. A study also suggested the connection between solar system's crossing of our Galaxy's spiral arm with Ice ages due to the increase in cosmic ray activity [25]. However, new calculations with improved map of the Galaxy have disproved this hypothesis [26].

## 2.2. Ionization and atmospheric chemistry

In the process of cosmic ray induced atmospheric ionization, the triple bond of $N_2$ and the double bond of $O_2$ are broken, which results in combination of the two species in a variety of ways and results in changes in atmospheric chemistry [27]. A quantitative estimate of such changes can be made using a photochemical model. It takes on an average 35 eV in the atmosphere to generate an ion pair [28]. Energy deposition by cosmic rays in the atmosphere can be computed using an air shower simulator such as CORSIKA [29], and the number of ion pairs can be computed. Lookup tables are also available which can be used to calculate ion pairs corresponding to the spectrum of any astrophysical source [15, 30]. This input can then be used in an atmospheric photochemical model to compute the corresponding changes in atmospheric chemistry.

Changes in atmospheric chemistry have significant implications due to the presence of the ozone layer in the upper atmosphere. The ozone layer is known to block the harmful UVB radiation, which directly interacts with the DNA, causing damage. Here are some representative reactions occurring in the upper atmosphere:

$$NO + O_3 \rightarrow NO_2 + O_2$$
$$NO_2 + O \rightarrow NO + O_2$$

The net reaction is: $O_3 + O \rightarrow O_2 + O_2$

A detailed description of all the reactions can be found elsewhere [31]. However, since high-energy primaries deposit most of their energy lower in the atmosphere, the effect on the ozone layer does not scale directly with the primary energy [27].

Other than ozone depletion and production of ozone at lower altitude, a number of oxides of nitrogen, or $NO_x$ species are also produced. Nitrates can be deposited on the ground through rain and act as fertilizers. This can increase plant growth on short timescales. The main damage due to ozone depletion is by the passage of solar UVB (290-315 nm), which is strongly absorbed by the DNA and protein molecules. This damaging effect is very important, especially for simple organisms such as phytoplankton, which form the base of the food chain and are responsible for half of the world's oxygen production [32]. It can also impact the growth of higher plant life and damage the skin of animals [14].

Secondary electrons also reach the ground, but they are of much lower energy and are much less penetrating than muons. Electrons can be stopped by 2 cm of water and therefore have limited effect on marine life.

**2.3. Radiation from secondary muons**

Once a muon is produced from the decay of daughter pions, which in turn are produced by parent hadrons in the upper atmosphere, its physics is relatively simple. Since its hadronic interaction cross section is extremely small, it undergoes energy loss primarily due to ionization. On an average, a muon loses 2 MeV/gcm$^2$ of energy in the atmosphere. A typical muon reaching the ground loses around 2 GeV in the atmosphere in form of ionization. The mean energy of a muon at the ground is around 4 GeV and therefore has a capability to penetrate the upper earth's crust as well as several hundred meters in water [33]. Since much of the biosphere is confined to this region, muons pose a significant threat. Muons can interact with the DNA and lead to mutations and cancerous diseases in organisms [34]. More work on muon damage is needed, because they are a poorly researched, significant form of radiation damage but subdominant under normal conditions. They are not increased by such events as nuclear reactor accidents, so there has been very little motivation to study their effects. But under cosmic ray excursions, they will dominate the direct dose on the ground. In some cases, such as a nearby supernova, they are expected to comprise a lethal dose.

Unlike other forms of radiation, damage caused by muons is approximately independent of the muon energy. This is because the energy deposition caused due to ionization (dE/dx) is a very slow function of energy. Therefore, at any given altitude, the flux of muons is more important than its energy for evaluating the biological damage. The energy of the muon becomes important in the case of the sub surface biosphere. Higher energy particles can easily damage organisms several hundreds of meters below the surface. Organisms living under rocks and inside caves, which are well shielded from other forms of radiation such as UV, are still subject to damage from muons.

Muon fluxes from cosmic rays can also be modeled using CORSIKA with high accuracy. Lookup tables are available using which one can take the cosmic ray spectrum from any source and calculate the muon flux on ground [33].

**2.4.   Radiation from neutrons**

Neutrons are produced in spallation reaction by cosmic ray primaries of energies greater than 1 GeV. Since they are electrically neutral particles, they do not lose energy by means of electromagnetic interactions, like other particles, but lose energy through collisions (short-range strong interactions).

Some neutrons are involved in reactions called neutron capture, which results in isotope formation and serves as a good proxy to track changes in cosmic ray flux. Typical cosmic ray primaries result in a peak of neutron flux in the stratosphere [35]. Even with high-energy primaries, there is no significant radiation dose from neutrons on ground. Due to their large cross sections, neutrons pose a significant threat in the upper atmosphere, especially at airline altitude. Lookup tables are available to compute the neutron flux for cosmic ray spectrum of any astrophysical source[35].

### 3. Signatures

Since a number of isotopes are produced by spallation reaction from cosmic ray primaries, a variation in the rate of cosmic rays can be determined from the measurement of concentration of these isotopes. $^{14}$C data is used extensively for this purpose. Since the half-life of $^{14}$C is 5730 years, it serves as a useful isotope to track cosmic ray variations over 10-50,000 year timescales. For larger timescales, its concentration becomes too low to draw any reasonable conclusion about the rate of cosmic rays. For variations over longer timescales, other geoisotopes need to be considered. $^{10}$Be has a half-life of 1.6 million years and is suitable for a few million-year timescales. $^{26}$Al has a half-life of 717,000 years. However, it is not possible to trace back cosmic ray variations over longer timescales and the only approach is to make reasonable assumptions based on rates of supernovae etc. in the galaxy. There are reports of higher than normal abundance of live $^{60}$Fe (half life = 1.5 Myr) deposits in deep ocean ferromanganese crust. The measured abundance is significantly higher than what one expects from terrestrial background. Based on calculations, it has been suggested that a nearby supernova (~ 30pc, ~ 5Myr) event might be responsible for such a signature [12]. This hypothesis can be tested by searching for a similar signal in $^{10}$Be, $^{129}$I, $^{146}$Sm and $^{53}$Mn.

Another approach is to measure the concentration of nitrates in ice core samples, which is directly proportional to the amount of ionization in the atmosphere. As discussed earlier, an increase in atmospheric ionization caused by cosmic ray primaries would result in changes in atmospheric chemistry and lead to the formation of nitrates. These nitrates are then washed from the atmosphere by rain and get deposited on the ground. Ice core samples too can be used to track back ~ 100,000 year history. However, recent work has argued that nitrate concentration cannot be used to determine the statistics of solar events [36].

### 4. Cosmic rays and biological radiation dose

Damage caused by radiation is directly proportional to the amount of energy deposited in the irradiated substance. Absorbed radiation dose, D is defined as the amount of energy absorbed per unit mass of the substance, D = dE/dM. The standard unit of D is Gray, defined as 1J/kg. This quantity is a broad indicator of the level of damage and corresponding radiobiological and clinical effects irrespective of the type and nature of the radiation source.

Another quantity to characterize biological damage is LET (Linear Energy Transfer), which is the average amount of energy deposited per unit length of the substance or LET = dE/dx. LET can be computed using the Bethe-Bloch equation for any given radiation source and substance. This equation gives at a microscopic level the amount of ionization produced in a substance and is a finer indicator of radiation damage as compared to the radiation dose.

It must also be mentioned that cells have natural repair mechanisms and the resulting biological damage cannot be completely characterized by simple physics equations. Since organisms are very complex and heterogeneous, radiation damage at certain sites have greater impact compared to others. Also, different radiation types produce different types of ionization trails, which are energy dependent. For example, a 10 keV photon and a 10 keV alpha particle will have very different ionization tracks and will produce different biological effects. In other words, different radiation types produce different effects, even if the absorbed dose is same. Some radiation types are more effective than others and a different measure called the Relative Biological Effectiveness (RBE) is used to quantify such differences. RBE is determined experimentally for a particular biological system in a given set of conditions. Since environmental conditions such as temperature, pressure, pH, etc. also affect biological damage, these are kept constant in order to experimentally determine RBE values for a given sample. $^{60}$Co is widely used as the reference radiation in order to calculate RBE. Each radiation type is assigned a weighting factor, $w_R$ after experimentation, which gives its RBE. The Sievert (= 1J/kg) is the SI unit of measuring the effective radiation dose incorporating RBE. Although Gray is also calculated in J/kg, Sievert also includes the weighting factor accounting for RBE.

The total annual radiation dose at the Earth's surface from all natural sources is 2.4 mSv/yr. A large fraction of this is due to naturally occurring radioisotopes. Major damage is caused to lungs by inhaling radon, which is present in the atmosphere and emits an alpha particle. Cosmic rays contribute a small fraction (0.39 mSv/yr annually, 85% from muons [34]). This level varies with the altitude and latitude because of atmospheric and geomagnetic effects. Excursions in cosmic ray intensity can bring large increases in muon dose, up to hundreds of mSv/yr for a nearby supernova.

## 5. Biological Implications

### 5.1. Origin of Life

The origin of life is one of the biggest challenges of modern science. In 1953, Miller and Urey's experiment demonstrated that simple amino acids could be produced by inducing electric discharges in a mixture of gases believed to be present in the prebiotic Earth [37]. As discussed earlier, an increase in the cosmic ray flux can increase the rate of lightning and therefore can contribute to the possible origin of life. However, models suggest that the Archean Earth experienced a cosmic ray flux reduced by two orders of magnitude or more lower than the present value [38]. This is due to enhanced solar activity, faster rotation and different structure of the Parker spiral, which was far more effective in shielding GCRs in the early history of the planet. For the present era, it has been shown that there is a high probability of increase in the rate of high energy cosmic rays over 100 Myr window [11]. Also, movement of the solar system in the Galaxy could lead to order of magnitude changes in cosmic ray intensity due to large differences in the density of interstellar medium [38]. A combination of these effects could have drastically increased the lightning rate and hence the production of building blocks of life.

## 5.2. Evolution of life

Mutations are caused when energetic photons or particles cause damage to the DNA molecule. Only ionizing radiation is capable of causing such damage, therefore most sunlight is not destructive. (However, UV, which is mostly shielded by the atmosphere, can be dangerous, particularly if doses are increased). This makes evolutionary sense too, because life could not have existed in presence of background radiation with too high a mutation rate. However, as we have discussed earlier, astrophysical objects affecting the Earth can cause a radiation burst, which in turn can trigger the mutation process on a large scale.

When it comes to visible and near-visible photons, the most damaging radiation is UV, specifically UVB and UVC. UVC is very effectively shielded by the atmosphere. UVB is easily absorbed by the DNA and causes maximum damage. In order to get an increase in UVB the incident radiation should have enough energy to ionize the upper atmosphere and deplete the ozone layer. Once the ozone layer is depleted, solar UVB can easily pass through it and directly interact with the DNA. Sources of such radiation can be galactic gamma-ray bursts (GRBs), nearby supernovae, or energetic solar proton events.

Galactic GRBs, although not frequent, have a non-trivial probability of impacting the Earth on timescales of the order of a few hundred million years. The energy released by GRBs is up to $10^{45}$ J within a few seconds to a minute. Short GRBs have less total energy, but a harder spectrum as compared to the long, and are more frequent and therefore, total energy likely to be deposited in the atmosphere from such an event is similar. It has also been shown that the dominant parameter affecting the atmospheric damage is the total fluence, followed by the spectrum [39]. The time development of the ionizing event has very little effect.

Energetic solar flares occur much closer to the Earth at a higher rate but the energy in a given event is several orders of magnitude smaller. It has been shown in some cases that particles are accelerated to an energy of ~20 GeV in extreme cases. The rate is not well constrained for large events [40]. In such cases, in addition to the damage caused to the atmosphere, muons will be produced from extreme hard spectra events, causing damage to the surface and near-surface organisms. Since the energy of such particles is high enough to penetrate the geomagnetic field, the extent of damage will be global, unlike low energy particles, which are mostly concentrated near polar regions.

On timescales of a few hundred million-years, nearby supernovae can significantly increase the flux of high-energy cosmic rays [11]. This can significantly increase the effects of solar UVB and muon radiation dose on the surface [34, 41].

Periodic variations of the order of ~60 Myr may increase the flux of cosmic rays to ~ PeV levels. The increase in rate is expected to last for a few million years. The total increase in radiation dose from muons is very important in this case. The total radiation dose enhancement is estimated to be between 1.26 – 4.36 times the total annual radiation dose from natural sources [34]. This mechanism could also explain the observed~60 Myr periodicity during the ~0.5 Gy period of a good fossil record [10, 34, 42-44].

## 6. Discussion

In the late 20s, John Joly suggested the possibility of cosmic radiation impacting living organisms and the long term evolution of life [45]. He also did pioneering work studying the links between radioactive sources and cancer rates [46]. Victor Hess and others realized that secondary particles produced by cosmic ray interactions can induce biological effects in addition to atmospheric ionization and primary radiation itself [47]. As the knowledge of the production sources of cosmic rays grew, ideas of the impacts of cosmic rays from nearly supernovae emerged [48-50]. Along with supernovae, the effects of high-energy protons were also estimated [51]. Along with theoretical estimates, experimental work was conducted, especially to study the effects during space travel [52-54].

We have presented a brief account of both periodic and non-periodic sources of cosmic ray variations over long timescales and their effects on terrestrial life. Signatures of cosmic ray variability have negligible half-life compared to the timescales considered here. This leaves us with no option other than to make reasonable assumptions about different cosmic ray sources and estimate their terrestrial effects. These assumptions will improve with our understanding of the production mechanisms of high and ultra-high energy cosmic rays and properties of their sources. Terrestrial effects include damage from solar UVB through ozone depletion, secondary muons and lightning discharge. Radiation doses from various astrophysical sources can be calculated and their biological effects can be estimated. Better experiments, especially with GeV muons are needed to get insights into its mechanism of damage on a variety of samples. The effects of radiation on various living organisms are very different, depending on their complexity and need to be explored further. A significantly bigger challenge is in translating this biological damage to the biosphere at large and in estimating its effects on the evolution of life.

"One can best feel in dealing with living things how primitive physics still is."
Albert Einstein

**References**

1. Dartnell, L.R., *Ionizing Radiation and Life.* Astrobiology, 2011. **11**(6): p. 551-582.
2. Karam, P.A., *Inconstant sun: How solar evolution has affected cosmic and ultraviolet radiation exposure over the history of life on earth.* Health Physics, 2003. **84**(3): p. 322-333.
3. Melott, A.L. and B.C. Thomas, *Astrophysical Ionizing Radiation and Earth: A Brief Review and Census of Intermittent Intense Sources.* Astrobiology, 2011. **11**(4): p. 343-361.
4. Melott, A.L. and B.C. Thomas, *Causes of an AD 774-775 C-14 increase.* Nature, 2012. **491**(7426): p. E1-E2.
5. Reames, D.V., *Particle acceleration at the Sun and in the heliosphere.* Space Science Reviews, 1999. **90**(3-4): p. 413-491.
6. Tammann, G.A., W. Loffler, and A. Schroder, *The Galactic Supernova Rate.* Astrophysical Journal Supplement Series, 1994. **92**(2): p. 487-493.
7. Aharonian, F.A., et al., *High-energy particle acceleration in the shell of a supernova remnant.* Nature, 2004. **432**(7013): p. 75-77.
8. Waddingt.Cj, *Paleomagnetic Field Reversals and Cosmic Radiation.* Science, 1967. **158**(3803): p. 913-&.
9. Bennett, K.D., *Milankovitch Cycles and Their Effects on Species in Ecological and Evolutionary Time.* Paleobiology, 1990. **16**(1): p. 11-21.
10. Medvedev, M.V. and A.L. Melott, *Do extragalactic cosmic rays induce cycles in fossil diversity?* Astrophysical Journal, 2007. **664**(2): p. 879-889.
11. Erlykin, A.D. and A.W. Wolfendale, *Long Term Time Variability of Cosmic Rays and Possible Relevance to the Development of Life on Earth.* Surveys in Geophysics, 2010. **31**(4): p. 383-398.
12. Fields, B.D. and J. Ellis, *On deep-ocean Fe-60 as a fossil of a near-earth supernova.* New Astronomy, 1999. **4**(6): p. 419-430.
13. Pavlov, A.A., et al., *Catastrophic ozone loss during passage of the Solar system through an interstellar cloud.* Geophysical Research Letters, 2005. **32**(1).
14. Thomas, B.C., A.L. Melott, and K.R. Arkenberg, *Terrestrial effects due to possible astrophysical sources of an AD 774-775 increase in $^{14}C$ production* submitted, 2012.
15. Atri, D., A.L. Melott, and B.C. Thomas, *Lookup tables to compute high energy cosmic ray induced atmospheric ionization and changes in atmospheric chemistry.* Journal of Cosmology and Astroparticle Physics, 2010(5).
16. Gurevich, A.V., et al., *Kinetic theory of runaway breakdown in inhomogeneous thundercloud electric field.* Physics Letters A, 2001. **282**(3): p. 180-185.

17. Gurevich, A.V. and K.P. Zybin, *Runaway breakdown and electric discharges in thunderstorms.* Physics-Uspekhi, 2001. **44**(11): p. 1119-1140.
18. Tinsley, B.A., *The global atmospheric electric circuit and its effects on cloud microphysics.* Reports on Progress in Physics, 2008. **71**(6).
19. Alexeenko, V.V., et al., *Transient variations of secondary cosmic rays due to atmospheric electric field and evidence for pre-lightning particle acceleration.* Physics Letters A, 2002. **301**(3-4): p. 299-306.
20. Svensmark, H., *Influence of cosmic rays on Earth's climate.* Physical Review Letters, 1998. **81**(22): p. 5027-5030.
21. Erlykin, A., T. Sloan, and A. Wolfendale, *Cosmic rays and climate.* Physics World, 2011. **24**(10): p. 25-25.
22. Kirkby, J., et al., *Role of sulphuric acid, ammonia and galactic cosmic rays in atmospheric aerosol nucleation.* Nature, 2011. **476**(7361): p. 429-U77.
23. Kasting, J.F., *Methane and climate during the Precambrian era.* Precambrian Research, 2005. **137**(3-4): p. 119-129.
24. Shaviv, N.J., *Toward a solution to the early faint Sun paradox: A lower cosmic ray flux from a stronger solar wind.* Journal of Geophysical Research-Space Physics, 2003. **108**(A12).
25. Shaviv, N.J., *The spiral structure of the Milky Way, cosmic rays, and ice age epochs on Earth.* New Astronomy, 2003. **8**(1): p. 39-77.
26. Overholt, A.C., A.L. Melott, and M. Pohl, *Testing the Link between Terrestrial Climate Change and Galactic Spiral-Arm Transit (Vol 705, Pg L101, 2009).* Astrophysical Journal Letters, 2012. **751**(2).
27. Melott, A.L., et al., *Atmospheric consequences of cosmic ray variability in the extragalactic shock model: 2. Revised ionization levels and their consequences.* Journal of Geophysical Research-Planets, 2010. **115**.
28. Porter, H.S., C.H. Jackman, and A.E.S. Green, *Efficiencies for Production of Atomic Nitrogen and Oxygen by Relativistic Proton Impact in Air.* Journal of Chemical Physics, 1976. **65**(1): p. 154-167.
29. Heck, D., et al., *CORSIKA: A Monte Carlo Code to Simulate Extensive Air Showers.* 1998. **6019**.
30. Usoskin, I.G. and G.A. Kovaltsov, *Cosmic ray induced ionization in the atmosphere: Full modeling and practical applications.* Journal of Geophysical Research-Atmospheres, 2006. **111**(D21).
31. Thomas, B.C., et al., *Terrestrial ozone depletion due to a milky way gamma-ray burst.* Astrophysical Journal, 2005. **622**(2): p. L153-L156.
32. Thomas, B.C. and M.D. Honeyman, *Amphibian Nitrate Stress as an Additional Terrestrial Threat from Astrophysical Ionizing Radiation Events?* Astrobiology, 2008. **8**(4): p. 731-733.
33. Atri, D. and A.L. Melott, *Modeling high-energy cosmic ray induced terrestrial muon flux: A lookup table.* Radiation Physics and Chemistry, 2011. **80**(6): p. 701-703.
34. Atri, D. and A.L. Melott, *Biological implications of high-energy cosmic ray induced muon flux in the extragalactic shock model.* Geophysical Research Letters, 2011. **38**.
35. Overholt, A.C., A.L. Melott, and D. Atri, *Modeling cosmic ray proton induced terrestrial neutron flux: A lookup table, submitted.*

36. Wolff, E.W., et al., *The Carrington event not observed in most ice core nitrate records.* Geophysical Research Letters, 2012. **39**.
37. Miller, S.L., *A Production of Amino Acids under Possible Primitive Earth Conditions.* Science, 1953. **117**(3046): p. 528-529.
38. Cohen, O., J.J. Drake, and J. Kota, *The cosmic ray intensity near the archean earth.* The Astrophysical Journal, 2012. **760**(85).
39. Ejzak, L.M., et al., *Terrestrial consequences of spectral and temporal variability in ionizing photon events.* Astrophysical Journal, 2007. **654**(1): p. 373-384.
40. Melott, A.L. and B.C. Thomas, *Causes of an AD 774-775 14C increase.* Nature, 2012.
41. Atri, D., *Terrestrial Effects of High-Energy Cosmic Rays.* Ph. D. Dissertation, University of Kansas, 2011.
42. Rohde, R.A. and R.A. Muller, *Cycles in fossil diversity.* Nature, 2005. **434**(7030): p. 208-210.
43. Melott, A.L. and R.K. Bambach, *A ubiquitous similar to 62-Myr periodic fluctuation superimposed on general trends in fossil biodiversity. I. Documentation.* Paleobiology, 2011. **37**(1): p. 92-112.
44. Melott, A.L. and R.K. Bambach, *A ubiquitous similar to 62-Myr periodic fluctuation superimposed on general trends in fossil biodiversity. II. Evolutionary dynamics associated with periodic fluctuation in marine diversity.* Paleobiology, 2011. **37**(3): p. 383-408.
45. Joly, J., *Cosmic radiations and evolution.* Nature, 1929. **123**: p. 981-981.
46. Joly, J., *Radium and cancer.* British Medical Journal, 1929. **1929**(1): p. 42-42.
47. Hess, V.F., J. Eugster, and P. Scherrer, *Cosmic radiation and its biological effects.* Fordham University Press, 1949.
48. Laster, H., W.H. Tucker, and K.D. Terry, *Cosmic Rays from Nearby Supernovae - Biological Effects.* Science, 1968. **160**(3832): p. 1138-&.
49. Ruderman, M.A., *Possible Consequences of Nearby Supernova Explosions for Atmospheric Ozone and Terrestrial Life.* Science, 1974. **184**(4141): p. 1079-1081.
50. Gehrels, N., et al., *Ozone depletion from nearby supernovae.* Astrophysical Journal, 2003. **585**(2): p. 1169-1176.
51. Wdowczyk, J. and A.W. Wolfendale, *Cosmic-Rays and Ancient Catastrophes.* Nature, 1977. **268**(5620): p. 510-512.
52. Kranz, A.R., et al., *Biological Damage Induced by Ionizing Cosmic-Rays in Dry Arabidopsis-Seeds.* Nuclear Tracks and Radiation Measurements, 1990. **17**(2): p. 155-165.
53. Cucinotta, F.A. and M. Durante, *Cancer risk from exposure to galactic cosmic rays: implications for space exploration by human beings.* Lancet Oncology, 2006. **7**(5): p. 431-435.
54. Kim, M.H.Y., K.A. George, and F.A. Cucinotta, *Evaluation of skin cancer risk for lunar and Mars missions.* Space Life Sciences: Flight Measurements, Calibration of Detectors and Environmental Models for Radiation Analysis, 2006. **37**(9): p. 1798-1803.